# GOD AND THE BIG-BANG: PAST AND MODERN DEBATES BETWEEN SCIENCE AND THEOLOGY

**Gabriele Gionti, S.J.**

Specola Vaticana (Vatican Observatory)

V-00120 Vatican City, Vatican City State and

Vatican Observatory Research Group

Steward Observatory, the University of Arizona

933 N. Cherry Ave, Tucson AZ, USA.

National Laboratories of Frascati (LNF) of Italian Institute of Nuclear Physics (INFN)

Via Enrico Fermi 54, 00044 Frascati (Rome), Italy.

## Abstract

A short phenomenological account of the genesis and evolution of the universe is presented with an emphasis on its primordial phases as well as its physical composition, i.e., dark Matter and dark Energy. We discuss Einstein's theory of General Relativity and its consequences for the birth of modern relativistic astrophysics. We introduce the Big Bang theory of Mons Lemaître as well as the competing theory of a Steady State Universe of Fred Hoyle. Since Big Bang theory appeared quite in agreement with Christian doctrine of creation, Pope Pius XII delivered a message to the Pontifical Academy of Sciences in 1951 claiming a certain agreement between the creation account in the book of Genesis and the Big-Bang theory (a concordist view), a position which he did not repeat later. On the other hand, Lemaître always kept separate the scientific and theological planes as two parallel "lines" never intersecting, i.e, as two complementary *magisteria*. Similar kind of tensions, between science and theology, emerge also today with the Hartle-Hawking solution to the Wheeler-DeWitt equation in quantum cosmology and its related speculations. To avoid some sort of confusion between theological and physics concepts, we, briefly, summarize the concept of creation in Christian theology.

1.

## 2. Introduction: a brief history of the Universe's evolution

The standard model of cosmology predicts that our universe began with a very dense and hot phase of matter and energy that has been called the "Big-Bang"[1], with the density and temperature of this initial phase approaching infinity. With time (between one and three minutes), this hot "plasma" universe began to expand and cool, reaching the temperatures necessary to form the nuclei of the first atoms: hydrogen and helium with traces of deuterium. Evidence of this primordial nucleo-synthesis can be gauged from the fact that helium accounts for nearly 24% of the weight of the baryonic matter in the Universe, an amount which could not be produced only in stars. The photons of the early radiation could not escape because they were scattered by the free electrons. When the temperature of the universe cooled down to 4000 K[2], the hydrogen atoms are able to recombine and the energy radiation of the hot gas is allowed to escape. This primordial radiation was emitted around 380.000 years after the Big-Bang, and which can be detected even today as the Cosmic Microwave Background Radiation (CMB). This primordial radiation in the universe bears the imprint of its last scattering from the ionized matter. The CMB today is homogeneous and isotropic and has a constant temperature except for very tiny fluctuations. The fact that the CMB has these tiny variations in temperature implies that the ionized gas, from which CMB originated, had very small density fluctuations. These tiny density fluctuations, under the effect of the gravitational field, collapsed and formed the first structures in the universe, galaxies, and clusters of galaxies. The formation of structures happened around one million years after the Big-Bang.

As regards the fundamental interactions in the history of the universe, at the beginning of the universe, between 0 and $10^{-43}$ sec, that is the Planck time at a temperature above $10^{32}$ K, Quantum

---

[1] William Stoeger, "The Big-Bang, quantum cosmology and *creation ex nihilo*" in *Creation and the God of Abraham*, ed. by David B. Burrell, Carlo Cogliati, Janet M. Soskice and William R. Stoeger (Cambridge: Cambridge University Press, 2010), 152-175.
[2] Stoeger, "The Big-Bang, quantum cosmology and *creation ex nihilo*", 155.

Mechanics and the Gravitational field were unified, most probably, in a unique theory called Quantum Gravity. Many scientists believed that quantization of gravity happened by the unification of all fundamental forces, including gravity, into one fundamental interaction. Although this is plausible, it has not been proved yet. It could be, pairwise, that only gravity was quantized in the Planck Era. Space and time were quantized as well. The gravitational interaction is the only interaction that exists at Planck scales and very high temperature, all other interactions will disappear. This is a possible scenario as well. Above the Planck Era, the gravitational field was present but not quantized, space and time and all other fundamental interactions, unified in one fundamental interaction called GUT (Grand Unified Theory)[3], emerged. The GUT is a quantized theory. This phase of the universe happened around a temperature below $10^{32}$ K. The GUT is the quantum unification of the strong, weak, and electromagnetic interactions into one unified interaction. This unification too is far to have been experimentally proven yet, but it looks plausible to many theoretical particle physicists. While the universe expanded and cooled down, we reached a temperature of $10^{26}$ K. The strong interaction separated from the electro-weak interaction. Before reaching this temperature, the Universe underwent a period of very rapid expansion, an exponential growth, called cosmological inflation. Cosmological inflation was caused by Dark Energy, an ingredient of the Universe of which we will talk later. Around $10^{15}$ K the electromagnetic interaction decoupled from weak interaction. The expansion and cooling of the Universe from the Planck era to $10^{15}$ K lasted one minutes. From one to three minutes the Universe behaved like a gas, as we have already highlighted above.

## 2. General relativity and the birth of relativistic cosmology

The Universe expands according to the laws of Einstein's General Relativity. This theory of gravitation was elaborated by Albert Einstein. Albert Einstein is rightly considered the scientist who, more than any other, made fundamental contributions to physics in the 20th century. He tried to formulate a theory of gravity that was a field theory and in which there was no action at a distance, as in

---

[3] Stoeger, "The Big-Bang, quantum cosmology and *creation ex nihilo*", 156.

Newton's theory. For Newton, a massive body "immediately" feels the presence of another body, and it can be said that this system implies a propagation with an infinite speed of gravitational perturbations. Obviously, Einstein knew Maxwell's theory of electromagnetism well and wondered how gravity could be described, not as an action at a distance, but as a field whose perturbations propagate with the speed of light. It took him 10 years from the discovery of special relativity to answer this question, and finally, in 1915, he came to the formulation of general relativity. This theory represents, in the history of physics, the beginning of a marriage between physical theories and complex mathematical theories. In fact, general relativity would not exist without Riemannian geometry, or rather without Lorentzian geometry[4]. The theory of general relativity is based on one fundamental postulates. This postulate says that the gravitational mass of each body is equal to its inertial mass, that is, the numerical value of the mass by which two bodies attract each other is equal to that of the inertial mass, which indicates how a body opposes movement. A "corollary" of it is the principle of covariance, according to which the laws of physics are the same, therefore covariant, in every reference system. In particular, this theory includes non-inertial reference systems, i.e., those that have a relative acceleration with respect to each other. (In the theory of special relativity, reference systems that have only a constant relative speed with respect to each other are taken into account) In this way space-time becomes a physical entity, which is no longer indifferent to physical phenomena, but is modified by the presence of massive bodies or the presence of energy and acquires a curvature. So, gravity is no longer a force at a distance, but becomes a field theory[5]. This means that, if I have a body of mass $m_1$ and move its position, another body of mass $m_2$ will feel the displacement (perturbation) of the position of $m_1$ not immediately, but after a time equal to the time it takes the light to travel the distance that separates the bodies $m_1$ and $m_2$. A consequence of all this is that if a ray of light that is emitted from a distant star to reach us passes close to the sun, it is deflected by the curvature generated by the mass of the sun, so that its apparent position with respect to an observer on earth does not coincide with its actual position.

---

[4] Steven Weinberg, *Gravitation and Cosmology*, (New York: John Wiley and Son, 1972), 19.

[5] Steven Weinberg, *Gravitation and Cosmology*, 67-70.

Immediately after the publication of the theory of general relativity, many physicists and mathematicians tried to derive exact solutions from the equations to which it gave rise. Friedmann, Lemaître, Robertson and Walker (FLRW)[6], independently of each other, found that, if we assume that the distribution of matter in the universe is homogeneous and isotropic, on a large scale, the solutions of the equations of general relativity foresee a universe that is, in the spatial part, a three-dimensional surface of a four-dimensional sphere, whose radius represents time. This sphere representing the Universe expands in time. To express this by analogy, three-dimensional space behaves as if it were a two-dimensional spherical surface on which all the galaxies and elements in general of the universe are located. Like a fairground balloon, this sphere expands, so the distance between galaxies increases over time.

## 3. The birth of Big-Bang cosmology and the concordism in theology

Einstein did not like this solution, and he branded it as "mathematically" correct but "physically" wrong. For this reason, he modified the equations of general relativity, introducing a constant, called "the cosmological constant," which provided a solution of a static universe that did not expand[7]. However, the measurement of the redshift of the spectral lines of the *nebulae* (galaxies) by Hubble proved the recession of all galaxies from an observer on Earth, demonstrating that the universe was actually expanding. When he realized this, Einstein declared that he had made the biggest mistake of his life.
But if the universe expands, then, going back in time, there must have been a primordial period when it was very small. This gave rise to the idea of the Belgian priest and cosmologist Georges Lemaître, who hypothesized that in the beginning the universe was the size of an atom (which he called "the original atom"), and that therefore the laws governing this original universe-atom were those of quantum mechanics.

---

[6] Steven Weinberg, *Gravitation and Cosmology*, 407-409.

[7] Steven Weinberg, *Gravitation and Cosmology*, 613-616.

Lemaître also had the distinction of having deduced the existence of the recession of the galaxies from the cosmological model FLRW, purely theoretically, before Hubble's measurement. However, as there were still no accurate measures of the velocities and photometric distances of the galaxies, he published his article in a little-known French scientific journal[8], so the credit for the famous law that bears his name went to Hubble[9].

This view of the evolution of the universe aroused much suspicion among many scientists, who noticed a close proximity to the biblical episode of creation in the Book of Genesis. To make fun of Lemaître's theory, Fred Hoyle, an English astrophysicist, called this theory the "Big Bang." He developed his own theory called the Steady State Universe, in which the universe expanded while maintaining a constant density of energy-matter, so that it had no beginning and no end, but it was necessary to assume a continuous production of matter-energy[10].

These two models of the universe remained in competition with each other for several years. On November 22, 1951, Pius XII – who was certainly one of the pontiffs most attentive to scientific questions – delivered a speech at the Pontifical Academy of Sciences. It was entitled "Un'ora"(A hour)[11] and in it he hinted that the cosmological model of the Big Bang confirmed the story of the creation of the world in the Book of Genesis. Of course, Pius XII was a very intelligent person and highlighted that more philosophical as well as theological investigations were needed to arrive to the conclusion that the Big-Bang theory and the *Fiat Lux* in Genesis were the same thing. But he was convinced, as he wrote it in his speech, that an enlightened mind could have made a jump (*a Pindaric flight*) and recognized traces of the God of creation and Love in the scientific description of the beginning and evolution of the Universe.

---

[8] International Astronomical Union, Thirtieth General Assembly: Resolution Presented to the XXXth General Assembly, Resolution B4, October 29 2018, https://www.iau.org/static/archives/announcements/pdf/ann18029e.pdf.

[9] Dominique Lambert, *The Atom of the Universe* (Krakow: Copernicus Center Press, 2016) 121-145. Two years ago, the International Union of Astronomy recognized Lemaître's merit and established that Hubble's Law can be called Hubble-Lemaître's Law. Elisabeth Gibney, "Belgian priest recognized in Hubble-law name change" *Nature News*, October 30, 2018 https://www.nature.com/articles/d41586-018-07234-y .

[10] Steven Weinberg, *Gravitation and Cosmology*, 459-464.

[11] Pope Pius XII, "Un'ora", November 22, 1951, https://www.vatican.va/content/pius-xii/it/speeches/1951/documents/hf_p-xii_spe_19511122_di-serena.html

In this speech, with its clearly Neo-Thomistic approach, the pope re-proposed the "ways" of establishing the existence of God of Saint Thomas Aquinas, especially the first and the fifth, based, respectively, on mutability and finality. In the Neo-Thomistic vein, he brought, to support the mutability (St. Thomas used the terms "generation" and "corruption" to indicate this mutability in the world), the processes of continuous change observed in physical and chemical phenomena, and to support the finality he brought into play the second principle of thermodynamics, according to which in the processes of nature the entropy of a closed physical system always increases. Therefore, increase of entropy, in the physical world, was evidence of a teleology. This theological approach, in which scientific theories were used to confirm theological positions, was later renamed "concordism."

Lemaître felt called into question by this speech, because in the past he had already been suspected of concordism. Moreover, a problem then arose, because the following year the meeting of the International Astronomical Union (IAU) was to be held in Rome, and Pius XII had been invited to give the inaugural speech. Therefore, Lemaître left South Africa to go to Rome, where, through the mediation of the Jesuit Fr. O'Connell, the then director of the Vatican Observatory, he met Pius XII. Obviously, we do not know the content of the conversation between the pope and Lemaître. The fact is that Pius XII gave his inaugural speech at the IAU on September 7, 1952, but made no mention of concordism[12].

For his part, Lemaître continued to always keep the theological and scientific planes distinct, as two parallel planes that do not intersect or, better still as two independent sources of knowledge. In 1965 two scientists from Bell Laboratories, Penzias, and Wilson, thanks to a large microwave antenna built for communication, detected a uniform radiation in all directions with a temperature of about 3 degrees Kelvin. This radiation, now known as the "Cosmic Microwave Background" (CMB), represents the first light emitted by the universe 400,000 years after the Big Bang, and can only be explained by the Big Bang theory, and not by that of the "Steady State Universe[13]."

---

[12] Józéf Turek, "Georges Lemaître and the Pontifical Academy of Sciences", *Vatican Observatory Publications* 2, no 13, (1989) 167-175.
[13] Steven Weinberg, *Gravitation and Cosmology*, 511.

Today the scientific community agrees that the universe in which we live was born 13.82 billion years ago, from a very hot phase, involving a cosmological event we call the "Big Bang." In the initial moment, called "the singularity," Einstein's equations are no longer valid. Immediately afterward, the universe underwent a great expansion, at a much greater rate than it is expanding now, an exponential expansion known as "inflation." About 400,000 years after the Big Bang, the universe emitted its first light, and then, little by little, all the structures were formed. In 1998, the study of the redshift of the light spectrum from Type IA supernovas showed that the universe was not only expanding, but also accelerating. Now, if the force responsible for this expansion is gravity alone, then the universe should expand by decelerating. If it accelerates, it means that a force opposite to gravity, a sort of anti-gravity, is operating. To explain this acceleration, the cosmological constant that Einstein had introduced in his equations was taken up again and the hypothesis of the existence of a non-visible energy called "dark energy" was formulated. In this way one gets a system that explains an accelerated expansion of the universe[14].

The nature of "dark energy" is not yet clear, and moreover it has not yet been directly observed. According to the latest measurements provided by the Planck satellite, "dark energy" should account for 68.3% of all energy-matter in the universe[15]. To this "exotic" element from the point of view of observations is also added "dark matter." In fact, the rotation curves of spiral galaxies have a radial velocity graph as a function of the distance from the center of the galaxy that does not coincide with the theoretical graph, which can be explained by the presence within the galaxy of non-conventional matter, which is called "dark matter". The latter is 26.8% of the total matter-energy of the universe, while the matter-energy observed in the universe is only 4.9%. It is therefore understood that this model of the universe, called $\Lambda$CDM ($\Lambda$ is the cosmological constant and refers to dark energy; CDM stands for *Cold*

---

[14] Steven Weinberg, *Cosmology*, (Oxford, Oxford University Press, 2008), 1-100.

[15] Planck Collaboration, "Planck 2018 cosmological parameters (Corrigendum)", *Astronomy & Astrophysics*, 652, C14 (August 6, 2021):1-3, https://doi.org/10.1051/0004-6361/201833910e.

*Dark Matter*, not high-energy dark matter), has many aspects that are still research topics and that do not allow us to say that a definitive model has been established[16].

## 4.Quantum gravity and some questions involving science and faith

Now we need to delve into the "Quantum Gravity" phase of our universe; what Lemaître had called the "primitive atom," because it has given rise to many debates on questions of science and faith.

As it is usually classified, quantum gravity is a phase of our universe that goes from the initial instant to Planck time, which is about $10^{-43}$ s. It is a very small interval of time, in which Einstein's equations, which we mentioned above, lose their predictive meaning. Therefore, we need a new theory that combines two worlds of physics that seem irreconcilable: quantum mechanics, which provides the laws of physics for the behavior of particles at the atomic and subatomic levels, and Einstein's general relativity, which describes the behavior of bodies on very large scales, beyond the galactic scales. This theory, which "should" – the conditional here is necessary, since we do not yet have a definitive theory – combine general relativity and quantum mechanics is called "quantum gravity."

One of the first approaches to this theory is the so-called "canonical approach," which basically consists in the attempt to write an equation for the wave function that should indicate the entire primordial universe. This is Wheeler-DeWitt's equation[17] and is without the variable "time," so it is said that the wave function of the primordial universe is timeless. This has generated a lot of confusion; however, it should be noted that a parameter of evolution is still necessary to describe the evolution of the universe:

---

[16] Planck Collaboration, "Planck 2015 results XIV. Dark Energy and Modified Gravity", *Astronomy and Astrophysic,* 594 A14 (September 20 2016):1-31, https://doi.org/10.1051/0004-6361/201525814.
[17] Edward W. Kolb and Michael S. Turner, *The Early Universe,* (New York: Addison Wesley, 1994), 447-464.

for example, in some cases, as the universe expands, the volume of the universe is used as a parameter of evolution.

Hartle and Hawking have come up with a solution for the Wheeler-DeWitt equation that goes under the name "Hartle-Hawking proposal." It is a fairly complex solution from a mathematical point of view, which aims to eliminate the problem of the initial "singularity." The Hartle-Hawking model suggests a sort of phase transition, in Planck time, from the Lorentz regime to the Riemannian regime. Thus, under Planck time, there are compact surfaces that have no singularity, and therefore no privileged points. For this reason, as Hawking has repeated in public lectures and in several of his writings, there is no beginning and there is no need for a God who acts as a "first cause" to initiate the process through which the universe evolves. Hartle and Hawking claim that, under Planck time, time is imaginary and therefore behaves like other spatial coordinates. The transition from phase to Planck time makes the transition from imaginary time to the physical time of true evolution. This "Riemannian phase" of the universe, which goes from the initial instant of the universe to Planck time, is the "Vacuum state" of the Hartle-Hawking model.

Hawking maintains that the Vacuum state is the *nihil* of the doctrine of *creatio ex nihilo*, and that the imaginary time under Planck time would explain the absence of time "required" by *creatio ex nihilo*. As the Jesuit William Stoeger[18] has pointed out, excessive manipulation is being carried out here by Hawking. The *nihil* in the above mentioned doctrine means really "nothing", the absence of everything even of energy. Nothing existed; even the physical laws were absent. While, in reality, in this quantum vacuum, there exists both energy and the physical laws that regulate phenomena. Moreover, to say that time does not exist in the sub-Planckian region, because time is imaginary, is also far-fetched.

The problem is that the Big Bang and "singularity" refer to an original event whose cause is unknown, and scientists fear that this cause must be a God who, like a demiurge, gives birth to the universe and then disappears, as suggested by some forms of deism. For this reason, Hawking felt the need to develop a model of quantum gravity that is completely autonomous and does not need to resort to an original cause: that is, according to him, one can do without God.

---

[18] William R. Stoeger, "Is Big-Bang Cosmology in Conflict with Creation?, in *The Heavens Proclaim*, ed. Guy Consolmagno (Vatican City: , Vatican Observatory Publication, 2009) 174-181.

However, there are two points to be clarified here. The first is that the Hartle-Hawking model is not *the* fundamental solution to quantum gravity, but *a* possible solution, which we do not even know whether it has occurred in nature[19]. The second is that to think that one has to resort to a God-demiurge to explain a cause one cannot clarify otherwise is to commit a philosophical error. Descartes made a similar error when he resorted to the existence of a good God to make sure that no one (the famous argument of Descarte's devil) had deceived him when he was building his philosophical system. This God who is used when something cannot be explained is called by some the "*God of the gaps.*"

But this is not a proper way of thinking in theology. In fact, if one day it were discovered that there is a phase of the universe before the Big Bang – and pre-Big Bang theories already exist – then this God-demiurge would no longer be useful, because science would explain that there is something before the Big Bang, and therefore God would not "useful" anymore and then it does not exist.

However, the problem of the beginning of the universe continues to be present in the minds of many scientists, because it is considered as a "prime cause" that needs recourse to a God-demiurge, especially since this beginning is confused with the term "creation."

The Christian concept of creation is, instead, completely different from that of the God-demiurge of scientists. First, God creates from a state where before there was really nothing (*creatio ex nihilo*), i.e., neither initial energy nor physical laws. Indeed, he creates both energy and physical laws from nothing and keeps them in existence. God creates the world and all its creatures into *being*. Creating creatures into *being*, he reduces to zero the distance between Him and them. He supports his own creation sustaining the creatures continuously into being (*creatio continua*). Creation is then a "relationship," as Saint Thomas Aquinas said or *creatio est relatio*[20], between the Father and the Son, who is the "Logos" through whom the Father creates the world and thanks to whom creation has a "logical" structure. This relationship between the Father and the Son is a relationship of Love, that is, the Holy Spirit, the third Person of the Trinity. In this way we have the *creatio ex amore*[21], so in creation we find the traces of God's Love. Therefore, creation has a basically Trinitarian structure[22]. There is a complementarity between Big-Bang cosmology and *creatio ex nihilo*. Big-Bang cosmology addresses, naturally, a question on creation that it cannot deal inside scientific categories. The doctrine of *creatio ex nihilo* answers, from a catholic theological perspective, the question on creation. Big-Bang cosmology and *creatio ex nihilo* can live together without conflicts if we accept, following Lemaître intuition, they are two different and distinct *magisteria*.

---

[19] William R. Stoeger, "Is Big-Bang Cosmology in Conflict with Creation?, 174-181.
[20] Thomas Aquinas, *Summa Theologiae*, I, q. 45, a.3.

## 5.Conclusions

In this essay we have summarized the temporal outline of our Universe. First, we have described the evolution of its physical elements starting from the Big-Bang up to the formation of its "cosmological" structures (Galaxies and clusters of Galaxies) in the Universe. Also, and in a parallel way, we have mentioned the history of the fundamental interactions in our Universe. We have introduced the Planck era, the Grand Unification theory (GUT), the strong and electroweak era up to nowadays four fundamental interactions. We have explained the basic elements of Einstein's General Relativity and modern relativistic cosmology. Then we have briefly mentioned that the FLRW homogeneous and isotropic cosmological model plus Dark Energy and Dark Matter constitute the $\Lambda$CDM standard model in cosmology. The history of the Big-Bang theory has been exposed with particular emphasis on Mons George Lemaître and the problem of "concordism" in theology. We have stressed that, after a "concordist" phase in his personal life, Lemaître always kept science and faith very distinct. Contemporary quantum cosmology model, as the Hartle-Hawking no-boundary boundary proposal, has been discussed. We have highlighted that any theological claim on the *creatio ex nihilo* based on theoretical physics considerations could be only completely misleading. We have briefly summarized what is *creatio ex nihilo* from pure theological considerations.

---

[21] Samuel Youngs, "Creatio Ex Amore Dei: Creation out of Nothing and God's Relational Nature", *The Asbury Journal* 69, 2 (2014): 165-186.

[22] Paolo Gamberini, *Un Dio relazione (*A relational God*)*, (Rome: Città Nuova, 2007) 148-163.

Today we see that the ΛCDM standard cosmological model works quite well with observational data; however, as we have explained, it is necessary to use ad hoc "elements," such as dark matter and dark energy, to explain some otherwise unexplained phenomena. In this sense one could think, with all the reservations and cautions of the case, that there could be an analogy between the theory of the epicycles of Ptolemy's geocentric system, invented to explain the motion of the planets, and the hypotheses of dark matter and dark energy, introduced to adapt the cosmological model to otherwise unexplained phenomena. In other words, it must be said that, despite all the progress that has been made in science, and in particular in current cosmology, the myth of a "very precise" science, without any shadow, must certainly be debunked. The truth, however, is that even the scientific models that we possess today and which we use to describe nature have limitations, and therefore do not possess to any degree the character of infallibility that a new dogmatic "scientism" would like to attribute to them.

Since ancient times there has always existed a tight connection between cosmology and religion. In ancient cultures, starting from the harmony and order existing in the visible universe – which at that time was simply the starry sky – people have always tried to hypothesize the existence of an "architect" God which is the cause of this harmony. Let us remember, one out of many, the so-called "cosmological proofs of the existence of God," where from the "contingency" of the world philosophical arguments deduced the necessity of the existence of a first cause, God, Who is also the guarantors of Universe harmony. However, the modern conflicts – for example, the "Galileo case" and the subsequent fracture between science and theology – lead us to think that, following Lemaître, the right approach, in the science-theology debate, is the separation between the theological and scientific planes or *magisteria*. But this does not prevent a mind, enlightened by the Grace of God as Pius XII was mentioning in his 1951 speech, from seeing in the harmony and order of the universe a beauty that reflects the imprint of the Creator and the Love with which God created and wove the universe. However, this is not proof of the existence of God, but rather an *a posteriori* observation, valid only for those who are either already believers or accepting God's Grace to believe.